\RequirePackage{lineno}
\documentclass[aps,prd,twocolumn,superscriptaddress,showpacs,amsmath,amssymb,nofootinbib]{revtex4-1}

\usepackage{comment}
\usepackage{multirow}
\usepackage{dcolumn}
\usepackage{bm}
\usepackage{setspace}
\usepackage{graphicx}
\includecomment{printrobustnotes}
\includecomment{printallnotes}
\usepackage{xspace}

\usepackage{dcolumn}
\usepackage{bm}

\newenvironment{cfigure1c}[1][tbp]{\begin{figure*}[#1]\centering}{\end{figure*}}

\newcommand{\note}[1]{\xspace}
\newcommand{\robustnote}[1]{\xspace}
\begin{printrobustnotes}
\renewcommand{\robustnote}[1]{\textbf{\textit{#1}}\xspace}
\end{printrobustnotes}
\begin{printallnotes}
\renewcommand{\note}[1]{\textbf{\textit{#1}}\xspace}
\renewcommand{\robustnote}[1]{\textbf{\textit{#1}}\xspace}
\end{printallnotes}
\newcommand{\ppbar}  {\ensuremath{p\bar{p}}\xspace}
\newcommand{\ttbar}  {\ensuremath{t\bar{t}}\xspace}

\newcommand{\Ht}     {\ensuremath{H_{T}}\xspace}
\newcommand{\et}     {\ensuremath{E_{T}}\xspace}
\newcommand{\etl}     {\ensuremath{E_{T}^{\ell}}\xspace}
\newcommand{\etj}     {\ensuremath{E_{T}^{\mathrm{jet}}}\xspace}

\newcommand{\pt}     {\ensuremath{p_{T}}\xspace}

\newcommand{\met}    {\ensuremath{\text{\raisebox{.3ex}{$\not$}}\et}\xspace}

\newcommand{\gmt}{\ensuremath{\Gamma_{\text{top}}}\xspace}
\newcommand{\gmtfit}{\ensuremath{\Gamma_{\text{meas}}}\xspace}

\newcommand{\mtop}    {\ensuremath{{M}_{\text{top}}}\xspace}

\newcommand{\mtreco}     {\ensuremath{m_{t}^{\text{reco}}}\xspace}

\newcommand{\mjj}    {\ensuremath{m_{\it jj}}\xspace}

\newcommand{\djes}   {\ensuremath{\Delta_{\mathrm{JES}}}\xspace}

\newcommand{\genunit}[2]{\ensuremath{#1~\mathrm{#2}}\xspace}

\newcommand{\gev}[1]    {\genunit{#1}{GeV}}
\newcommand{\pb}[1]     {\genunit{#1}{pb}}

\newcommand{\gevc}[1]   {\ensuremath{#1~\mathrm{GeV}/c}}
\newcommand{\gevcc}[1]  {\ensuremath{#1~\mathrm{GeV}/c^{2}}}

\newcommand{\invfb}[1]  {\ensuremath{#1~\mathrm{fb}^{-1}}}

\newcommand{\gevnoarg}  {\ensuremath{\mathrm{GeV}}\xspace}

\makeatletter
\AtBeginDocument{\@ifpackageloaded{natbib}{\ifNAT@numbers\if@filesw\immediate\write\@auxout{\string\global\string\NAT@numberstrue}\fi\fi}{}}
\makeatother
\begin{document}

\title{A direct measurement of the total decay width of the top quark}

\affiliation{Institute of Physics, Academia Sinica, Taipei, Taiwan 11529, Republic of China}
\affiliation{Argonne National Laboratory, Argonne, Illinois 60439, USA}
\affiliation{University of Athens, 157 71 Athens, Greece}
\affiliation{Institut de Fisica d'Altes Energies, ICREA, Universitat Autonoma de Barcelona, E-08193, Bellaterra (Barcelona), Spain}
\affiliation{Baylor University, Waco, Texas 76798, USA}
\affiliation{Istituto Nazionale di Fisica Nucleare Bologna, \ensuremath{^{jj}}University of Bologna, I-40127 Bologna, Italy}
\affiliation{University of California, Davis, Davis, California 95616, USA}
\affiliation{University of California, Los Angeles, Los Angeles, California 90024, USA}
\affiliation{Instituto de Fisica de Cantabria, CSIC-University of Cantabria, 39005 Santander, Spain}
\affiliation{Carnegie Mellon University, Pittsburgh, Pennsylvania 15213, USA}
\affiliation{Enrico Fermi Institute, University of Chicago, Chicago, Illinois 60637, USA}
\affiliation{Comenius University, 842 48 Bratislava, Slovakia; Institute of Experimental Physics, 040 01 Kosice, Slovakia}
\affiliation{Joint Institute for Nuclear Research, RU-141980 Dubna, Russia}
\affiliation{Duke University, Durham, North Carolina 27708, USA}
\affiliation{Fermi National Accelerator Laboratory, Batavia, Illinois 60510, USA}
\affiliation{University of Florida, Gainesville, Florida 32611, USA}
\affiliation{Laboratori Nazionali di Frascati, Istituto Nazionale di Fisica Nucleare, I-00044 Frascati, Italy}
\affiliation{University of Geneva, CH-1211 Geneva 4, Switzerland}
\affiliation{Glasgow University, Glasgow G12 8QQ, United Kingdom}
\affiliation{Harvard University, Cambridge, Massachusetts 02138, USA}
\affiliation{Division of High Energy Physics, Department of Physics, University of Helsinki, FIN-00014, Helsinki, Finland; Helsinki Institute of Physics, FIN-00014, Helsinki, Finland}
\affiliation{University of Illinois, Urbana, Illinois 61801, USA}
\affiliation{The Johns Hopkins University, Baltimore, Maryland 21218, USA}
\affiliation{Institut f\"{u}r Experimentelle Kernphysik, Karlsruhe Institute of Technology, D-76131 Karlsruhe, Germany}
\affiliation{Center for High Energy Physics: Kyungpook National University, Daegu 702-701, Korea; Seoul National University, Seoul 151-742, Korea; Sungkyunkwan University, Suwon 440-746, Korea; Korea Institute of Science and Technology Information, Daejeon 305-806, Korea; Chonnam National University, Gwangju 500-757, Korea; Chonbuk National University, Jeonju 561-756, Korea; Ewha Womans University, Seoul, 120-750, Korea}
\affiliation{Ernest Orlando Lawrence Berkeley National Laboratory, Berkeley, California 94720, USA}
\affiliation{University of Liverpool, Liverpool L69 7ZE, United Kingdom}
\affiliation{University College London, London WC1E 6BT, United Kingdom}
\affiliation{Centro de Investigaciones Energeticas Medioambientales y Tecnologicas, E-28040 Madrid, Spain}
\affiliation{Massachusetts Institute of Technology, Cambridge, Massachusetts 02139, USA}
\affiliation{University of Michigan, Ann Arbor, Michigan 48109, USA}
\affiliation{Michigan State University, East Lansing, Michigan 48824, USA}
\affiliation{Institution for Theoretical and Experimental Physics, ITEP, Moscow 117259, Russia}
\affiliation{University of New Mexico, Albuquerque, New Mexico 87131, USA}
\affiliation{The Ohio State University, Columbus, Ohio 43210, USA}
\affiliation{Okayama University, Okayama 700-8530, Japan}
\affiliation{Osaka City University, Osaka 558-8585, Japan}
\affiliation{University of Oxford, Oxford OX1 3RH, United Kingdom}
\affiliation{Istituto Nazionale di Fisica Nucleare, Sezione di Padova, \ensuremath{^{kk}}University of Padova, I-35131 Padova, Italy}
\affiliation{University of Pennsylvania, Philadelphia, Pennsylvania 19104, USA}
\affiliation{Istituto Nazionale di Fisica Nucleare Pisa, \ensuremath{^{ll}}University of Pisa, \ensuremath{^{mm}}University of Siena, \ensuremath{^{nn}}Scuola Normale Superiore, I-56127 Pisa, Italy, \ensuremath{^{oo}}INFN Pavia, I-27100 Pavia, Italy, \ensuremath{^{pp}}University of Pavia, I-27100 Pavia, Italy}
\affiliation{University of Pittsburgh, Pittsburgh, Pennsylvania 15260, USA}
\affiliation{Purdue University, West Lafayette, Indiana 47907, USA}
\affiliation{University of Rochester, Rochester, New York 14627, USA}
\affiliation{The Rockefeller University, New York, New York 10065, USA}
\affiliation{Istituto Nazionale di Fisica Nucleare, Sezione di Roma 1, \ensuremath{^{qq}}Sapienza Universit\`{a} di Roma, I-00185 Roma, Italy}
\affiliation{Mitchell Institute for Fundamental Physics and Astronomy, Texas A\&M University, College Station, Texas 77843, USA}
\affiliation{Istituto Nazionale di Fisica Nucleare Trieste, \ensuremath{^{rr}}Gruppo Collegato di Udine, \ensuremath{^{ss}}University of Udine, I-33100 Udine, Italy, \ensuremath{^{tt}}University of Trieste, I-34127 Trieste, Italy}
\affiliation{University of Tsukuba, Tsukuba, Ibaraki 305, Japan}
\affiliation{Tufts University, Medford, Massachusetts 02155, USA}
\affiliation{University of Virginia, Charlottesville, Virginia 22906, USA}
\affiliation{Waseda University, Tokyo 169, Japan}
\affiliation{Wayne State University, Detroit, Michigan 48201, USA}
\affiliation{University of Wisconsin, Madison, Wisconsin 53706, USA}
\affiliation{Yale University, New Haven, Connecticut 06520, USA}

\author{T.~Aaltonen}
\affiliation{Division of High Energy Physics, Department of Physics, University of Helsinki, FIN-00014, Helsinki, Finland; Helsinki Institute of Physics, FIN-00014, Helsinki, Finland}
\author{S.~Amerio\ensuremath{^{kk}}}
\affiliation{Istituto Nazionale di Fisica Nucleare, Sezione di Padova, \ensuremath{^{kk}}University of Padova, I-35131 Padova, Italy}
\author{D.~Amidei}
\affiliation{University of Michigan, Ann Arbor, Michigan 48109, USA}
\author{A.~Anastassov\ensuremath{^{w}}}
\affiliation{Fermi National Accelerator Laboratory, Batavia, Illinois 60510, USA}
\author{A.~Annovi}
\affiliation{Laboratori Nazionali di Frascati, Istituto Nazionale di Fisica Nucleare, I-00044 Frascati, Italy}
\author{J.~Antos}
\affiliation{Comenius University, 842 48 Bratislava, Slovakia; Institute of Experimental Physics, 040 01 Kosice, Slovakia}
\author{G.~Apollinari}
\affiliation{Fermi National Accelerator Laboratory, Batavia, Illinois 60510, USA}
\author{J.A.~Appel}
\affiliation{Fermi National Accelerator Laboratory, Batavia, Illinois 60510, USA}
\author{T.~Arisawa}
\affiliation{Waseda University, Tokyo 169, Japan}
\author{A.~Artikov}
\affiliation{Joint Institute for Nuclear Research, RU-141980 Dubna, Russia}
% ...

\collaboration{CDF Collaboration}
\altaffiliation[With visitors from]{
\ensuremath{^{a}}University of British Columbia, Vancouver, BC V6T 1Z1, Canada,
\ensuremath{^{b}}Istituto Nazionale di Fisica Nucleare, Sezione di Cagliari, 09042 Monserrato (Cagliari), Italy,
\ensuremath{^{c}}University of California Irvine, Irvine, CA 92697, USA,
\ensuremath{^{d}}Institute of Physics, Academy of Sciences of the Czech Republic, 182~21, Czech Republic,
\ensuremath{^{e}}CERN, CH-1211 Geneva, Switzerland,
\ensuremath{^{f}}Cornell University, Ithaca, NY 14853, USA,
\ensuremath{^{g}}University of Cyprus, Nicosia CY-1678, Cyprus,
\ensuremath{^{h}}Office of Science, U.S. Department of Energy, Washington, DC 20585, USA,
\ensuremath{^{i}}University College Dublin, Dublin 4, Ireland,
\ensuremath{^{j}}ETH, 8092 Z\"{u}rich, Switzerland,
\ensuremath{^{k}}University of Fukui, Fukui City, Fukui Prefecture, Japan 910-0017,
\ensuremath{^{l}}Universidad Iberoamericana, Lomas de Santa Fe, M\'{e}xico, C.P. 01219, Distrito Federal,
\ensuremath{^{m}}University of Iowa, Iowa City, IA 52242, USA,
\ensuremath{^{n}}Kinki University, Higashi-Osaka City, Japan 577-8502,
\ensuremath{^{o}}Kansas State University, Manhattan, KS 66506, USA,
\ensuremath{^{p}}Brookhaven National Laboratory, Upton, NY 11973, USA,
\ensuremath{^{q}}University of Manchester, Manchester M13 9PL, United Kingdom,
\ensuremath{^{r}}Queen Mary, University of London, London, E1 4NS, United Kingdom,
\ensuremath{^{s}}University of Melbourne, Victoria 3010, Australia,
\ensuremath{^{t}}Muons, Inc., Batavia, IL 60510, USA,
\ensuremath{^{u}}Nagasaki Institute of Applied Science, Nagasaki 851-0193, Japan,
\ensuremath{^{v}}National Research Nuclear University, Moscow 115409, Russia,
\ensuremath{^{w}}Northwestern University, Evanston, IL 60208, USA,
\ensuremath{^{x}}University of Notre Dame, Notre Dame, IN 46556, USA,
\ensuremath{^{y}}Universidad de Oviedo, E-33007 Oviedo, Spain,
\ensuremath{^{z}}CNRS-IN2P3, Paris, F-75205 France,
\ensuremath{^{aa}}Universidad Tecnica Federico Santa Maria, 110v Valparaiso, Chile,
\ensuremath{^{bb}}The University of Jordan, Amman 11942, Jordan,
\ensuremath{^{cc}}Universite catholique de Louvain, 1348 Louvain-La-Neuve, Belgium,
\ensuremath{^{dd}}University of Z\"{u}rich, 8006 Z\"{u}rich, Switzerland,
\ensuremath{^{ee}}Massachusetts General Hospital, Boston, MA 02114 USA,
\ensuremath{^{ff}}Harvard Medical School, Boston, MA 02114 USA,
\ensuremath{^{gg}}Hampton University, Hampton, VA 23668, USA,
\ensuremath{^{hh}}Los Alamos National Laboratory, Los Alamos, NM 87544, USA,
\ensuremath{^{ii}}Universit\`{a} degli Studi di Napoli Federico I, I-80138 Napoli, Italy
}
\noaffiliation

\date{\today}
\begin{abstract}
We present a measurement of the total decay width of the top quark using events with top-antitop-quark pair candidates reconstructed in the final state with one charged lepton and four or more hadronic jets. We use the full Tevatron Run~II data set of $\sqrt{s} = 1.96$~TeV proton-antiproton collisions recorded by the CDF II detector. 
The top-quark mass and the mass of the hadronically-decaying $W$ boson are reconstructed for each event and compared with distributions derived from simulated signal and background samples to extract the top-quark width~(\gmt) and the energy scale of the calorimeter jets with {\it in-situ} calibration. For a top-quark mass $\mtop = \gevcc{172.5}$, we find $1.10<\gmt<\gev{4.05}$ at 68\% confidence level, which is in agreement with the standard-model expectation of \gev{1.3} and is the most precise direct measurement of the top-quark width to date.
\end{abstract}

\maketitle

The top quark~($t$) is the heaviest known elementary particle. Its large mass endows it with the largest decay width, and hence, the shortest lifetime of any of the known fermions~\cite{pdg}. At leading order calculation of quantum chromodynamics~(QCD), the top-quark decay width~(\gmt) depends on the top-quark mass (\mtop), the Fermi coupling constant ($G_F$), and the magnitude of the top-to-bottom-quark coupling in the quark-mixing matrix ($|V_{tb}|$)~\cite{width_theory0}. 
The next-to-leading-order calculation with QCD and electroweak corrections predicts $\gmt = \gev{1.33}$ at $\mtop = \gevcc{172.5}$ with approximately 1\% precision~\cite{width_theory1,width_theory2}. This is consistent with the recent next-to-next-to-leading order calculation of $\gmt = \gev{1.32}$~\cite{width_nnlo}. A deviation from the standard-model~(SM) prediction could indicate the presence of non-SM decay channels, such as decays through a charged Higgs boson~\cite{chiggs}, the supersymmetric top-quark partner~\cite{susy}, or a flavor-changing neutral current~\cite{fcnc}. A direct measurement of \gmt provides  general constraints on such processes.

The D0 Collaboration has determined the width to be $\gmt = \gev{2.00^{+0.47}_{-0.43}}$ in a data set corresponding to an integrated luminosity of \invfb{5.4}, using a model-dependent, indirect measurement that assumes SM couplings~\cite{d0width}. 
The CDF Collaboration reported more model-independent measurements of the width using a direct shape comparison of the reconstructed top-quark mass in data to the simulated top-quark mass distributions~\cite{cdfwidth1,cdfwidth2}. The most recent measurement set an upper limit of $\gmt < \gev{7.6}$ at the 95\% confidence level~(C.L.) with a data set corresponding to \invfb{4.3}~\cite{cdfwidth2}. Even though the direct measurement is less precise  than the indirect one, it probes a broader class of non-SM physics models, because the direct measurement has less dependence on the SM. 

This paper reports on an direct measurement of the top-quark width in \ppbar collisions at the Tevatron, using the full Run~II data set, corresponding to an integrated luminosity of \invfb{8.7} collected with the CDF II detector~\cite{cdfdet1}, which is a general-purpose azimuthally and forward-backward symmetric detector surrounding the colliding beams of the Tevatron \ppbar collider.  
We not only increase statistical sensitivity using a larger sample with respect to Ref.~\cite{cdfwidth2}, but also improve jet-energy calibrations using an artificial neural network~\cite{nnjes}. 

Top quarks at the Tevatron are predominantly produced in $\ttbar$ pairs. We reconstruct top-quark decays in the topology of 
$t\rightarrow bW^{+}$ and $\bar{t}\rightarrow \bar{b}W^{-}$.
Events with a $W$ boson decaying into a charged lepton~(electron or muon) and a neutrino ($W\rightarrow {\ell}\nu$ including the cascade decay of $W \rightarrow \tau~(\rightarrow \ell \bar{\nu}) \nu$) and the other $W$ boson decaying into a pair of jets\footnote{Collimated sprays of particles resulting from the hadronization of quarks} defines the lepton~+~jets channel ($\ttbar \rightarrow \ell \nu b\bar{b} q\bar{q}$). 
To select $\ttbar$ candidate events in this channel, we require one electron (muon) with $\et > \gev{20}$ ($\pt > \gevc{20}$) and pseudorapidity $|\eta|<1.1$~\cite{cdfco}. 
We also require large missing transverse-energy~\cite{met} ($\met > \gev{20}$) and at least four hadronic jets. 
Jets are reconstructed by combining signals from particles detected within a spatial cone of radius $\Delta R = \sqrt{(\Delta \eta)^2+(\Delta \phi)^2} = 0.4$~\cite{jetclu}. 
Observed jet energies are corrected for nonuniformities of the calorimeter response parametrized as a function of $\eta$, the energy contributed by multiple \ppbar interactions in the event, and the calorimeter's nonlinear response~\cite{jes}. 
In addition to the standard jet-energy corrections, we use an artificial neural network that includes additional information, such as jet momentum from the charged particles inside the jet~\cite{nnjes}, to improve jet-energy resolution~\cite{ljmassfinal,massdiff}. 
Jets originating from $b$ quarks are identified~(tagged) using a secondary-vertex-tagging algorithm~\cite{secvtx}. 

We divide the sample of \ttbar candidates into subsamples with zero~(0-tag), one~(1-tag), and two or more~(2-tag) $b$-tagged jets, which have different signal-to-background ratios.
We further classify the events according to the jet kinematic properties. The ``tight" selection requires exactly four jets, each with $\et > \gev{20}$ and $|\eta|<2.0$. The ``loose" selection on the remaining events requires exactly three jets with  $\et > \gev{20}$ and $|\eta|<2.0$, and one or more additional jets with $\et>\gev{12}$ and $|\eta|<2.4$. We then combine the $b$-tag and jet-selection categories into five subsamples used in the analysis: 0-tagT, 1-tagL, 1-tagT, 2-tagL, and 2-tagT, where ``T" and ``L" denote the ``tight" and ``loose" jet selections.  
Finally, to reduce the level of non-\ttbar background contributions to the 0-tag and 1-tag samples, we require the scalar sum of transverse energies in the event, $\Ht = \etl + \met + \sum_{\text{four jets}}\etj$, to exceed 250~GeV. 

The primary sources of non-\ttbar backgrounds are $W$~+~jets and multijet production. We also consider small contributions from $Z$~+~jets, dibosons, and single-top quark production. 
The multijet background is estimated by the data-driven techniques described in Ref.~\cite{qcdback}. 
The kinematic distributions of $W$~+~jets are modeled with the {\sc alpgen}~\cite{alpgen} generator. The number of $W$~+~jets events is determined from the total number of events observed in data by subtraction of the expected \ttbar and the other backgrounds event contributions. 
Diboson backgrounds are modeled by {\sc alpgen} for ${\it WW}$, ${\it WZ}$, ${\it ZZ}$ and {\sc pythia}~\cite{pythia} for $W\gamma$, while single-top-quark processes are generated with {\sc madgraph}~\cite{madgraph}. We normalize simulated event yields using their theoretical next-to-leading-order cross sections~\cite{bgxsection}.  
References~\cite{secvtx,xsection} provide the details of these techniques. Table~\ref{table_background} summarizes the sample composition in each subsample.

To distinguish between different values of \gmt, we compare the reconstructed top-quark mass distribution observed in data to various distributions from \ttbar signal samples generated using {\sc pythia} with different \gmt values ranging from 0.1 to \gev{30} for a fixed $\mtop = \gevcc{172.5}$.
Because the jet energy scale~(JES) is one of the dominant systematic uncertainties in the analysis~\cite{cdfwidth1}, we generate a set of samples where the JES is varied independently. In the data, jet energies are corrected  to account for the energy scale error in the calorimeter with uncertainty, $\sigma_c$, the CDF JES fractional uncertainty~\cite{jes}. 
In the simulation, we vary the JES with the correction factor of jet energies, 1+\djes, with varing the values of \djes from $-3.0 \sigma_c$ to $+3.0 \sigma_c$.

\begin{table*}[ht]
\begin{center}
\caption[Expected number of background and signal events]{
Expected and observed numbers of signal and background events
assuming a \ttbar production cross section $\sigma_{\ttbar} = \pb{7.45}$ and $\mtop = \gevcc{172.5}$.
}
\begin{ruledtabular}
\label{table_background}
\begin{tabular}{lccccc}
             &0-tag       & 1-tagL     &1-tagT       &2-tagL      & 2-tagT\\
\hline
$W$~+~jets      &  703 $\pm$ 199  &170 $\pm$ 60 & 102 $\pm$ 37 & 11.6 $\pm$ 4.9 & 8.4 $\pm$ 3.5\\
$Z$~+~jets      & 52.3 $\pm$ 4.4 &8.9 $\pm$ 1.1& 5.9 $\pm$ 0.7&  0.8 $\pm$ 0.1 & 0.5 $\pm$ 0.1\\
Single top    &  4.8 $\pm$ 0.5  &10.5 $\pm$ 0.9&6.8 $\pm$ 0.6& 2.2 $\pm$ 0.3 & 1.7 $\pm$ 0.2 \\
Diboson       & 60.3 $\pm$ 5.6 &11.1 $\pm$ 1.4 &8.5 $\pm$ 1.1& 1.0 $\pm$ 0.2 & 0.8 $\pm$ 0.1\\
Multijets     & 143  $\pm$ 114  &34.5 $\pm$ 12.6& 20.7 $\pm$ 16.6&4.4 $\pm$ 2.5&2.5 $\pm$ 2.4\\
\hline
Background    & 963 $\pm$ 229& 235 $\pm$ 61& 144 $\pm$ 41 & 19.9 $\pm$ 5.5 &13.8 $\pm$ 4.2\\
\ttbar signal & 645 $\pm$ 86 &695 $\pm$ 87&867 $\pm$ 108&192 $\pm$ 30&304 $\pm$ 47\\
\hline
\hline
Expected      &1608 $\pm$ 245&930 $\pm$ 106&1011 $\pm$ 115&212 $\pm$ 30&318 $\pm$ 47\\
Observed      &  1627     &    882   &    997     & 208      & 275 \\
\end{tabular}
\end{ruledtabular}
\end{center}
\end{table*}

\begin{cfigure1c}
\begin{tabular}{cc}
\includegraphics[width=0.43\textwidth]{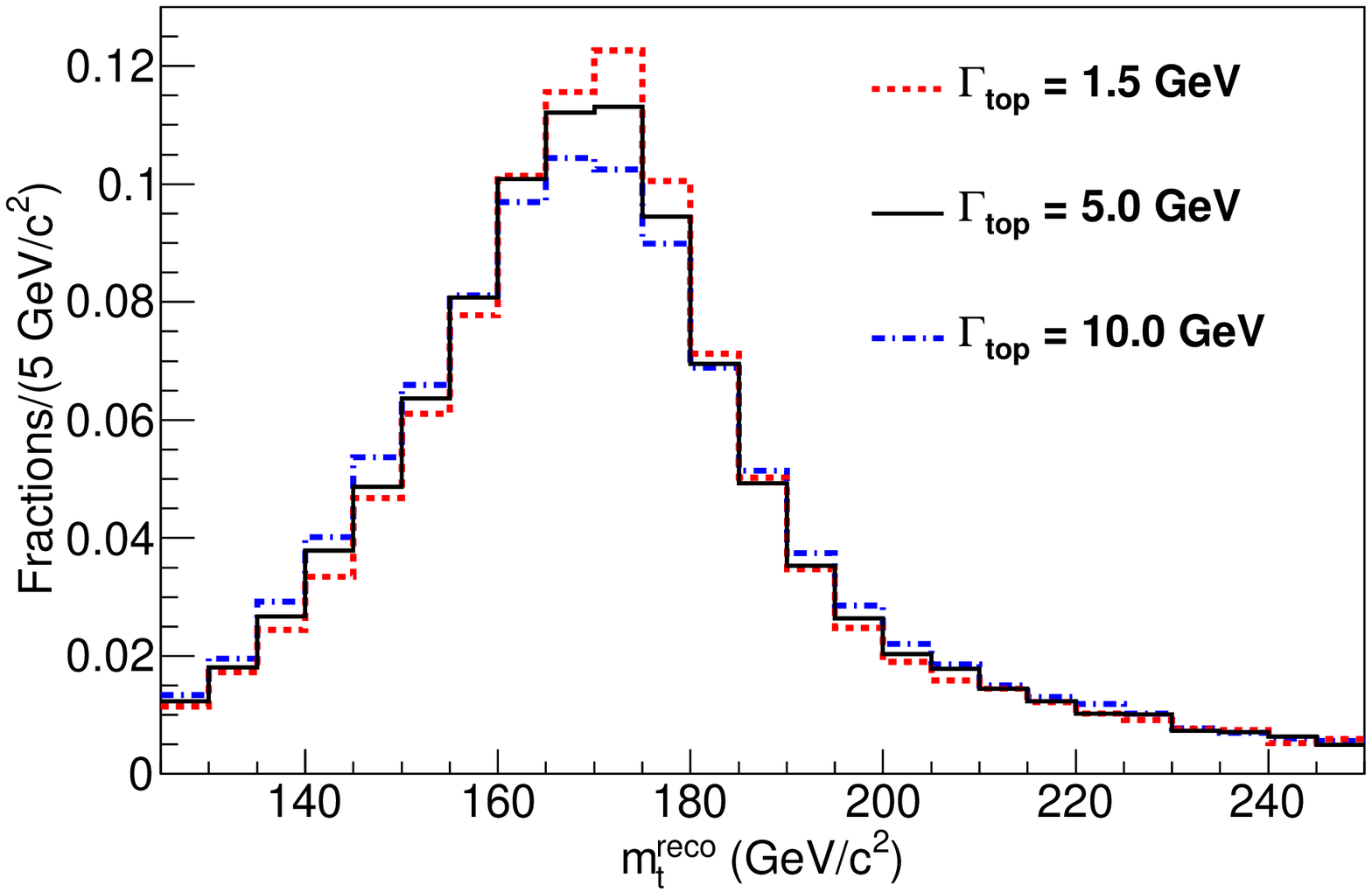} &
\includegraphics[width=0.43\textwidth]{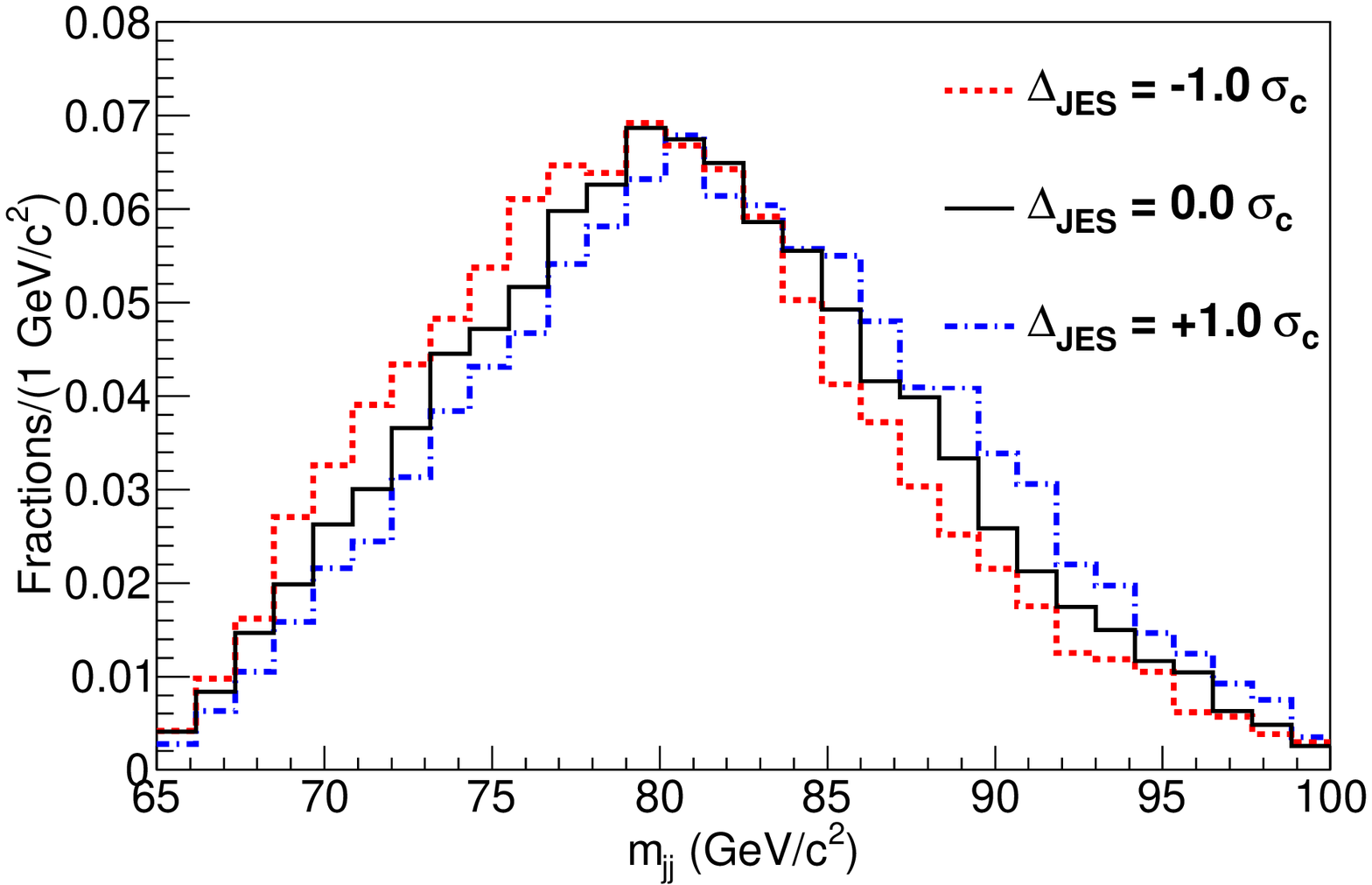} \\
(a) & (b) \\
\end{tabular}
\caption[Templates]{Distributions for simulated events meeting the lepton~+~jets selection: (a) \mtreco distributions  displayed with three values of \gmt and with the nominal $\djes = 0.0$; (b)  \mjj distributions  displayed with three values of \djes and with $\gmt = \gev{1.5}$.
}
\label{ref:template}
\end{cfigure1c}

After event selection, the analysis proceeds in three steps. First, we reconstruct a top-quark mass~(\mtreco), defined below, from each event. The width of the \mtreco distribution is a sensitive variable for \gmt. We also reconstruct the hadronically decaying $W$-boson mass~(\mjj). 
The constraint of \mjj to the known $W$-boson mass can be used to determine the JES calibration {\it in situ}, which reduces the dominant uncertainty from the JES. 
The second step is a likelihood fit of \mtreco and \mjj comparing with simulated signal and background distributions to determine \gmtfit, an estimator of \gmt, which will be explained later. Finally, we use a likelihood-ratio ordering to determine the 68\% and 95\% C.L. limits of \gmt from \gmtfit~\cite{fc}.

For the event reconstruction, we assume that all selected events are lepton~+~jets \ttbar events and perform a complete reconstruction of the \ttbar kinematic properties~\cite{tmtold,tmt19}.  
We perform a $\chi^2$ minimization to fit the momenta of the \ttbar decay products and determine \mtreco for each event using the four leading jets. To resolve the ambiguity arising from the jets-to-quarks assignments, we require that $b$-tagged jets are assigned to $b$ quarks and select the assignment with the lowest $\chi^2$. 
To reject events having  poorly-reconstructed kinematic properties, we request the minimum value of $\chi^2$ to be less than  9.0~(less than 3.0) for the $b$-tagged~(zero $b$-tag) events. The dijet mass, \mjj, is calculated independently as the invariant mass of two non-$b$-tagged jets that provides the closest value to the known $W$-boson mass, \gevcc{80.4}~\cite{wmass}.   Figure~\ref{ref:template}(a) shows the distributions of \mtreco for three different \gmt values. 
The shape of \mtreco depends on \gmt, yielding an estimate of its value. 
Distributions of \mjj for three different values of \djes are shown in Fig.~\ref{ref:template}(b). The maximum  of the distribution depends strongly on \djes. Hence, \mjj can be used to constrain the JES {\it in situ}.

To account for the correlation between \mtreco and \mjj, we construct two-dimensional p.d.f.s of signals and background with the two-dimensional kernel-density estimates~\cite{KDEHEP} for the likelihood fit procedure~\cite{tmt19}. First, at discrete values of \gmt from {0.1} to \gevcc{30} and \djes from $-3.0 \sigma_c$ to $+3.0 \sigma_c$, we estimate the p.d.f.s for the observables from the above-mentioned {\sc pythia} $\ttbar$ samples. Background p.d.f.s are estimated for various values of \djes from $-3.0 \sigma_c$ to $+3.0 \sigma_c$. We interpolate the simulated distributions to find p.d.f.s for arbitrary values of \gmt and \djes using a local polynomial smoothing method~\cite{lps}. Then, we fit the signal and background p.d.f.s to the unbinned distributions observed in data. In the fit of data, we apply a Gaussian constraint to the expected number of background events, but there is no constraints on the expected number of signal events. Separate likelihoods are constructed for the five subsamples, and the overall likelihood is obtained by multiplying them together. Maximization of the total likelihood yields the best-fit value \gmtfit. 

The limit on the true value of \gmt from the measured \gmtfit is set using the Neyman construction~\cite{neyman}. 
In this procedure, the unphysical region of negative \gmt is not allowed for \gmtfit, which makes acceptance region of \gmtfit to be equal or greater than zero. It makes the large number of events at \gmtfit equal to zero for a small \gmt. 
We derive the confidence bands from simulated experiments in which signal and background events are selected from the simulated samples.  

\begin{table}
\begin{center}
\caption[Summary of systematics]{Summary of systematic uncertainties on \gmt.}
\begin{ruledtabular}
\label{systtablesummary}
\begin{tabular}{lc}
Source              & Uncertainty~(\gevnoarg) \\
\hline
Jet resolution                     & 0.56 \\
Color reconnection                 & 0.69 \\
Event generator              & 0.50 \\
Higher-order effects               & 0.21 \\
Residual jet-energy scale          & 0.19 \\
Parton distribution functions      & 0.24 \\
$b$-jet energy scale               & 0.28 \\
Background shape                   & 0.18 \\
Gluon fusion fraction              & 0.26 \\
Initial- and final-state radiation  & 0.17 \\
Lepton energy scale                & 0.03 \\
Multiple hadron interaction        & 0.23 \\
\hline
Total systematic uncertainty       & 1.22 \\
\end{tabular}
\end{ruledtabular}
\end{center}
\end{table}

Because this measurement relies on the shape of \mtreco, the uncertainties on the JES calibration and the jet resolution could dominate. 
However, the JES  is well controlled with {\it in-situ} calibration using the \mjj distributions. 
To estimate the uncertainty from the jet-energy resolution, we use experimental and simulated data samples of events with a photon recoiling against a jet in the final state. In these samples, we estimate the energy of the jets using the energy of the recoiled photon. We compare the \pt-dependent resolutions on the energy of the reconstructed jets in data and simulation. We obtain consistent results within statistical uncertainty. Taking into account statistical uncertainty of the data, we define a \pt-depedent systematic uncertainty on jet resolution to cover the difference. 
In addition to the jet-energy resolution, the uncertainties associated with modeling of color flow in the interaction and with the arbitrary choice of the event generator are the dominant systematic uncertainties, as shown in Table~\ref{systtablesummary}.  The color-reconnection systematic uncertainty takes into account the effects of the underlying color structure of quarks and gluons and its flow~\cite{CR} by rearrangements from the simplest configuration to enhanced color reconnections based on simulations with differently-tuned configuration parameters~\cite{CR_tune}. For the systematic uncertainty associated with the choice of the event generator, the samples generated by {\sc pythia} and {\sc herwig}~\cite{herwig} are used. We examine the effects of higher-order corrections using {\sc mc@nlo}~\cite{mcnlo}, a full next-to-leading-order simulation. Other sources of systematic effects, including uncertainties in parton-distribution functions, initial- and final state gluon radiation, multiple hadron interactions, $b$-jet-energy scale, gluon fusion fraction, background shape, and lepton-energy scale, give small contributions. The total systematic uncertainty of \gev{1.22} is calculated as a quadrature sum of the listed uncertainties. We estimate the systematic uncertainties under the assumptions of $\mtop = \gevcc{172.5}$ and $\gmt = \gevcc{1.5}$, but checks with different values of \mtop and \gmt for the dominant sources show consistent results. The details of the systematic-uncertainty evaluations are described in Refs.~\cite{topmass_average,tmtold,tmt19}.

\begin{figure}
\includegraphics[width=0.45\textwidth]{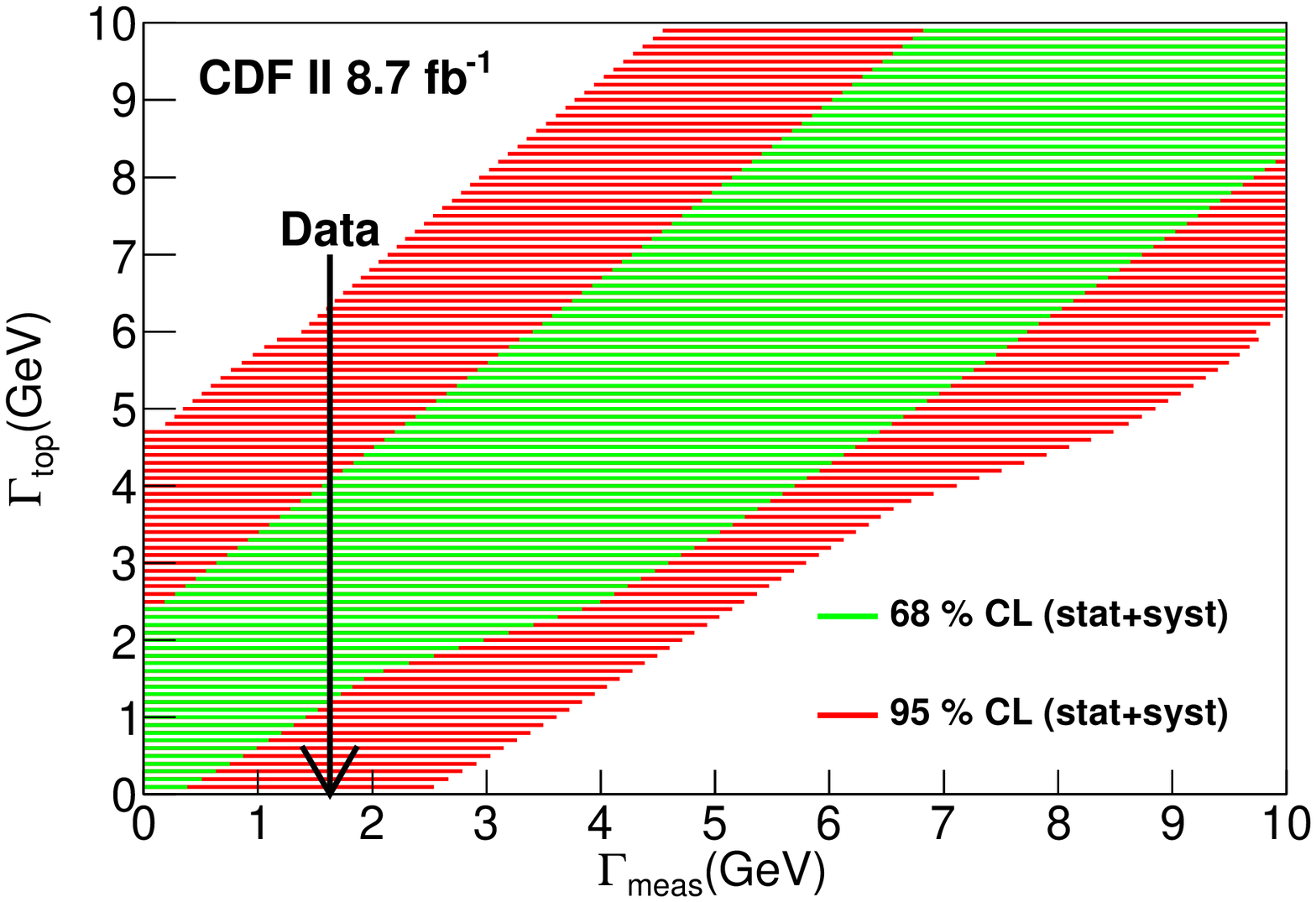}
\caption[FC]{Confidence bands of \gmt as a function of \gmtfit for  68\% and 95\% C.L. limits. Results from simulated experiments assuming \invfb{8.7} of data at different values of \gmt are convoluted with a smearing function to account for systematic uncertainties. The value observed in data is indicated by an arrow.
}
\label{ref:FC}
\end{figure}

To incorporate systematic effects into the confidence bands we use a convolution method for folding systematic effects into the likelihood function~\cite{conv_sys, conv_sys1} based on bayesian treatment of systematic uncertainties~\cite{sys_bay1,sys_bay2}. We convolve the likelihood function with a Gaussian p.d.f. that has a width equal to \gev{1.22} and is centered at zero.  We then build the confidence bands with 68\% and 95\% coverages as shown in Fig.~\ref{ref:FC}. The value of \gmtfit retrieved from the data is \gev{1.63} and is depicted as an arrow in the plot. This corresponds to an uppler limit of $ \gmt < \gev{6.38}$ at the 95\% C.L. We also set a two-sided limit of $\text{1.10}<\gmt<\gev{4.05}$ at the 68\% C.L., which corresponds to a lifetime of $1.6\times10^{-25} <\tau_{\text{top}}< 6.0\times10^{-25} \text{ s}$. For a typical quark hadronization time scale, $3.3\times10^{-24} \text{ s}$~\cite{qcdhad}, this result supports the assertion that top-quark decay occurs before hadronization.

In conclusion, a direct measurement of the top-quark width is performed in fully reconstructed lepton~+~jets events by using the full CDF Run II data set corresponding to an integrated luminosity of \invfb{8.7} of \ppbar collisions at $\sqrt{s}$~=~1.96~TeV. 
We obtain $\text{1.10}<\gmt<\gev{4.05}$ at 68\% C.L., which corresponds to a lifetime of $1.6\times10^{-25}<\tau_{\text{top}}< 6.0\times10^{-25} \text{ s}$.
This is the most precise direct determination of the top-quark width and lifetime and shows no evidence of non-SM physics in the top-quark decay.

\begin{acknowledgments}
	We thank the Fermilab staff and the technical staffs of the participating institutions for their vital contributions. This work was supported by the U.S. Department of Energy and National Science Foundation; the Italian Istituto Nazionale di Fisica Nucleare; the Ministry of Education, Culture, Sports, Science and Technology of Japan; the Natural Sciences and Engineering Research Council of Canada; the National Science Council of the Republic of China; the Swiss National Science Foundation; the A.P. Sloan Foundation; the Bundesministerium f\"ur Bildung und Forschung, Germany; the Korean World Class University Program, the National Research Foundation of Korea; the Science and Technology Facilities Council and the Royal Society, UK; the Russian Foundation for Basic Research; the Ministerio de Ciencia e Innovaci\'{o}n, and Programa Consolider-Ingenio 2010, Spain; the Slovak R\&D Agency; the Academy of Finland; the Australian Research Council (ARC); and the EU community Marie Curie Fellowship contract 302103. 
\end{acknowledgments}


\begin{thebibliography}{99}
\bibitem{cdfco} We use a right-handed spherical coordinate system with the origin at the center of the detector. The angles $\theta$ and $\phi$ are the polar and azimuthal angles, respectively, and $\theta$ is zero along incident proton direction.  The pseudorapidity is defined by $\eta = -\ln \tan(\theta/2)$.  The transverse momentum and energy are defined by $\pt = p \sin(\theta)$ and $\et = E \sin(\theta)$, respectively, where $p$ and $E$ are the momentum and energy of the particle.
\bibitem{met} The missing transverse energy, an imbalance of energy in the transverse plane of the detector, is defined by
$\met = -\vert \sum_{towers}\et{\hat n_{T}} \vert$, where ${\hat n_{T}}$ is the unit vector normal to the beam and pointing to a given calorimeter tower and $\et$ is the transverse energy measured in that tower.
\end{thebibliography}
\end{document}